\documentclass[twocolumn]{aastex631}

\usepackage{amsmath,amssymb,bm}

\newcommand{\msun}{\ensuremath{M_{\odot}}}
\newcommand{\mtwotwo}{\ensuremath{m_{22}}}
\newcommand{\Ms}{\ensuremath{M_{\rm s}}}
\newcommand{\Mc}{\ensuremath{M_{\rm c}}}
\newcommand{\rc}{\ensuremath{r_{\rm c}}}
\newcommand{\re}{\ensuremath{r_{\rm e}}}
\newcommand{\nh}{\ensuremath{N_{\rm H}}}

\newcommand{\safeincludegraphics}[2][]{%
  \IfFileExists{#2}{%
    \includegraphics[#1]{#2}%
  }{%
    \fbox{\parbox[c][0.25\columnwidth][c]{0.95\columnwidth}{\centering\textit{Figure File Missing: #2}}}%
  }%
}

\shorttitle{Born in the Dark: LRDs from FDM Solitons}
\shortauthors{Woo}

\begin{document}

\title{Born in the Dark: The Catastrophic Collapse of Fuzzy Dark Matter Solitons as the Origin of Little Red Dots}
\accepted{for publication in The Astrophysical Journal}

\correspondingauthor{Tak-Pong Woo}
\email{bonwood@phys.ntu.edu.tw}

\author[0009-0000-4728-6868]{Tak-Pong Woo}
\affiliation{Institute of Astrophysics, National Taiwan University, Taipei 10617, Taiwan}
\affiliation{Department of Physics, National Taiwan University, Taipei 10617, Taiwan}

\begin{abstract}
JWST surveys have uncovered a population of compact, red sources (``Little Red Dots,'' LRDs) at $z\ge 5$ that exhibit broad Balmer emission yet remain X-ray 
faint, implying heavy obscuration with $N_{\rm H}\ge 10^{24}\,{\rm cm^{-2}}$.
We propose that LRDs trace a short-lived, obscured phase associated with rapid baryonic inflow inside the deep solitonic cores of fuzzy dark matter (FDM) halos.
Combining the soliton size scaling with the observed compact radii ($r_e\sim 30$--$100$ pc), we find that while a particle mass of $m \sim 2 \times 10^{-22}$\,eV provides a direct match for mature systems, the observed size-mass relation is fully consistent with heavier bosons ($m_{22} \gtrsim 20$, satisfying Lyman-$\alpha$ constraints) if LRDs represent a \textit{progenitor phase} where the central black hole is still growing ($M_{\rm BH} < M_c$).
We adopt $m_{22}=2$ as a fiducial baseline to demonstrate the thermodynamic instability.
A conservative mass-budget estimate indicates that configurations reaching $N_{\rm H}\ge 10^{24}$--$10^{25}\,{\rm cm^{-2}}$ require densities for which radiative losses (cooling and/or diffusion) occur faster than the dynamical time, suggesting that a long-lived static hot atmosphere is unlikely (an ``Opacity Crisis'') and that rapid inflow or radiation-pressure-driven evolution is favored.
Using $512^3$ pseudo-spectral Schr\"odinger--Poisson simulations of idealized soliton mergers, we illustrate that compact, high-density soliton cores form robustly under representative scalings. We discuss observational implications and tests, and outline the need for future radiation-hydrodynamic modeling to predict demographics and detailed spectra.
\end{abstract}

\keywords{dark matter (353); early universe (432); quasars (1319); high-redshift galaxies (734); star formation (1569)}

\section{Introduction}\label{sec:intro}

The origin of supermassive black holes (SMBHs) residing in the centers of galaxies remains one of the most puzzling questions in modern astrophysics.
The detection of quasars at redshift $z > 7$ with black hole masses exceeding $10^9\,\msun$ \citep{Mortlock2011, Banados2018}---together with the growing set of JWST- and X-ray-selected 
massive black hole candidates at $z\gtrsim 4$--$10$ (e.g., \citealt{Yang2020, Larson2023, Kokorev2023, Kovacs2024, Bogdan2024, Bosman2024, Suh2025})---places stringent timing constraints on their growth.
Standard scenarios, such as light seeds from Pop III stars or heavy seeds from Direct Collapse Black Holes (DCBH), face significant challenges regarding accretion rates and rare environmental requirements \citep{Bromm2013}.
The advent of the James Webb Space Telescope (JWST) has added a new layer of complexity.
JWST surveys have revealed a ubiquitous population of ``Little Red Dots'' (LRDs) at $z \sim 5$--$10$ \citep{Labbe2023, Greene2024}.
These objects are characterized by their compact morphology, with effective (half-light) radii $r_e\lesssim 50$--$100$ pc, extremely red spectral energy distributions (SEDs), and broad H$\alpha$ lines indicating SMBH masses of $10^7$--$10^9\,\msun$.
Crucially, despite their apparent black hole activity, many LRDs are surprisingly X-ray faint \citep{Maiolino2024, Maiolino2025}, consistent with Compton-thick obscuration with column densities $\nh > 10^{24}$~cm$^{-2}$ \citep[e.g.,][]{Setton2025, Ji2025, InayoshiMaiolino2025}.
Fuzzy Dark Matter (FDM) provides a novel framework for addressing these issues.
Recent simulations have highlighted how FDM filaments can naturally reproduce the prolate morphology of high-redshift galaxies \citep{Pozo2025}.
Complementary to this large-scale picture, FDM halos host a stable, dense ground-state solution known as a \textit{soliton} \citep{Schive2014a}.
In this paper, we propose that LRDs are the direct observational signatures of the \textbf{catastrophic breakdown of the hydrostatic atmosphere} triggered by FDM soliton thermodynamics.
We identify a regime of ``Opacity Crisis'' where the static atmosphere becomes unphysical.
We present analytic derivations and 3D numerical simulations to show that: (1) deep solitonic potential wells form robustly, and (2) gas within these wells must undergo rapid cooling and inflow.
We organize the paper as follows. Section~\ref{sec:theory} develops the analytic framework connecting soliton scaling, achievable columns, and the timescale-based ``Opacity Crisis.''
Section~\ref{sec:radiative} discusses radiative implications and a parameterized clumpy cocoon geometry.
Sections~\ref{sec:method} and \ref{sec:results} describe our Schr\"odinger--Poisson simulations and demonstrate robust formation of compact soliton cores.
Section~\ref{sec:discussion} connects the model to Lyman-$\alpha$ constraints, size--mass trends, and demographic duty-cycle requirements, and Section~\ref{sec:conclusion} summarizes.

\section{Analytic Framework}\label{sec:theory}

\subsection{Soliton Scaling and Parameter Constraints}\label{sec:scaling}
The FDM soliton radius $\rc$ scales inversely with the boson mass $m$ and soliton mass $\Ms$ \citep{Schive2014a}:
\begin{equation}
    \rc \approx 160 \, \text{pc} \, \mtwotwo^{-2} \left(\frac{\Ms}{10^9 \msun}\right)^{-1},
\label{eq:rc_scaling}
\end{equation}
where $\mtwotwo = m / 10^{-22} \text{ eV}$.
We constrain $\mtwotwo$ by requiring consistency with LRD observables: compact sizes ($\re \sim 30$--$100$ pc) and Compton-thick obscuration ($\nh > 10^{24} \rm{cm}^{-2}$).
For a typical LRD host halo, the core--halo relation implies 
$\Ms \sim 10^9\,\msun$ \citep[e.g.,][]{Schive2014b, Chan2018}.

\begin{figure}[t]
    \centering
    \safeincludegraphics[width=\columnwidth]{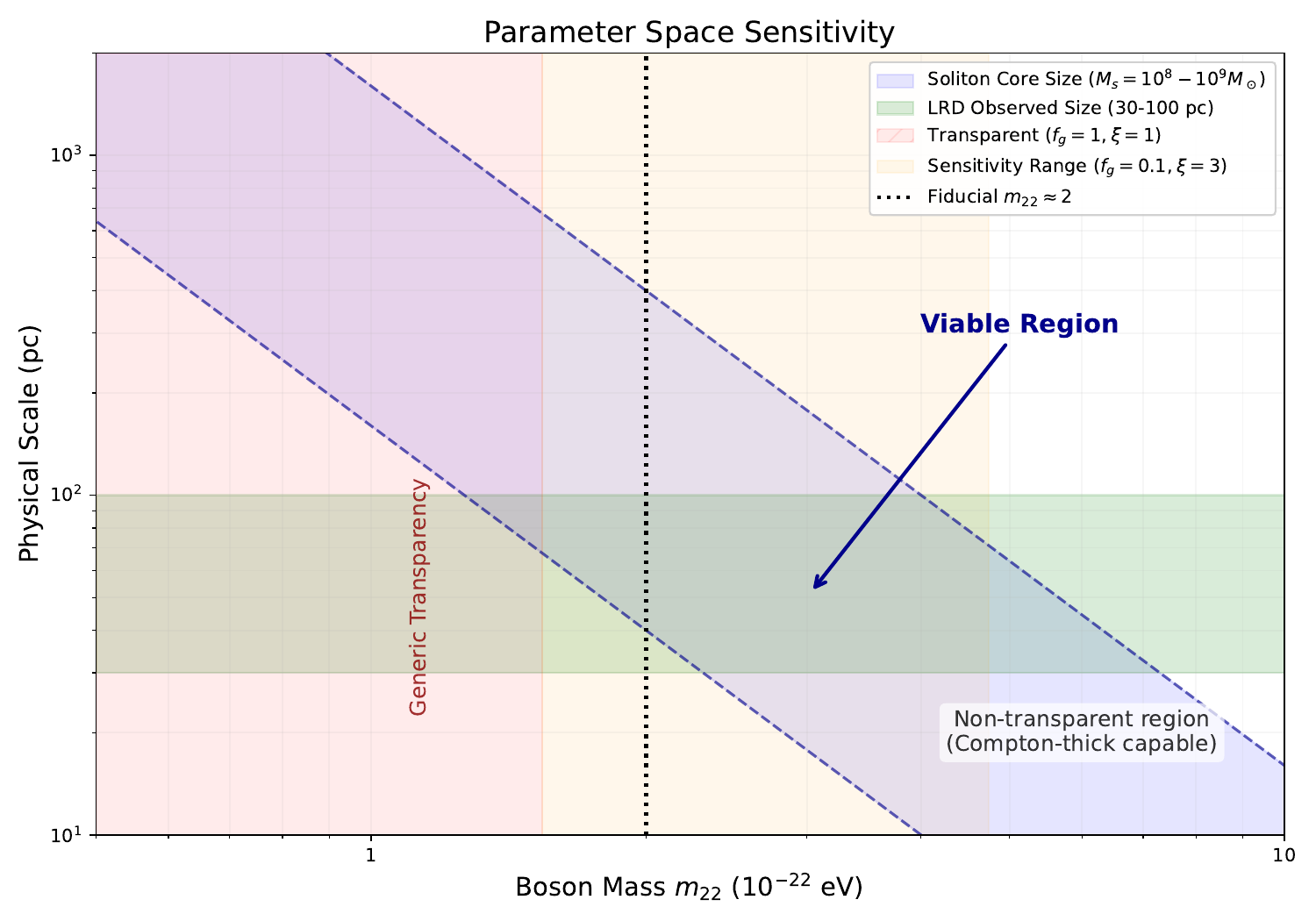}
    \caption{Constraints on boson mass ($m_{22}$) from combining the soliton size scaling (Eq.~\ref{eq:rc_scaling}) with an opacity (column) requirement.
The vertical axis shows a characteristic physical scale; we compare the observed effective radius range of LRDs (green band, $r_e\sim 30$--$100$~pc) with the soliton core radius predicted for $M_s=10^8$--$10^9\,M_\odot$ (blue band; dashed boundaries follow $r_c\propto m_{22}^{-2}$).
The red hatched region (``Too transparent''/``Generic transparency'') marks parameters for which even a maximally loaded core ($f_g=1$) confined within $r_{\rm e}=\xi r_c$ with $\xi=1$ cannot reach $N_{\rm H}=10^{24}\,{\rm cm^{-2}}$.
The opacity boundary is evaluated for a representative $M_s\sim 10^9\,M_\odot$ (typical of LRD hosts).
The orange shaded area illustrates sensitivity to baryon-loading uncertainties by adopting conservative values ($f_g=0.1$, $\xi=3$).
The viable region is the overlap between the observed size band and the non-transparent (Compton-thick-capable) region;
The region outside the red-shaded area corresponds entirely to the non-transparent (Compton-thick capable) regime.
The ``Viable Region'' is specifically the parallelogram defined by the overlap between the observed size band and this non-transparent regime.
under the direct-mapping assumption ($r_e\sim r_c$) this favors $m_{22}\sim{\rm few}$, and we adopt $m_{22}=2$ as a fiducial baseline for subsequent figures.}\label{fig:parameter_space}
\vspace{15pt}
\end{figure}

As shown in Figure~\ref{fig:parameter_space}, we identify the parameter space where FDM is consistent with observations.
To ensure robustness, we also perform a sensitivity analysis regarding the baryon loading parameters.
We note that the derived constraints depend on the assumption that the observed effective radius $r_e$ traces the soliton core scale $r_c$ (i.e., $\eta_r \equiv r_e/r_c \sim \mathcal{O}(1)$).
For the soliton scaling in Eq.~(\ref{eq:rc_scaling}), one finds $m_{22}\propto r_c^{-1/2}\propto \eta_r^{1/2}$ at fixed $(r_e, M_s)$;
thus an uncertainty $\eta_r\sim 0.5$--$3$ shifts the inferred $m_{22}$ by a factor of $\sim 0.7$--$1.7$.
If $r_e$ corresponds to a larger scattering photosphere, or if the black hole is in an early progenitor phase where it does not yet trace the full soliton mass ($M_{\rm BH} < M_c$), the constraints would shift significantly.
However, under the direct mapping assumption ($M_{\rm BH} \approx M_c$), a fiducial window around $m_{22} \approx 2$ emerges.

\subsection{Derivation of the Opacity Crisis}
We ask whether a quasi-static baryonic atmosphere inside a soliton can simultaneously (i) produce Compton-thick obscuration and (ii) remain pressure-supported on a dynamical time.
Rather than assuming a unique static solution, we explicitly parameterize the baryon loading and show that the regime required for obscuration generically implies rapid radiative losses, invalidating the long-lived hydrostatic picture.
We define the relevant dynamical timescale at radius $r$ as determined by the enclosed mass potential,
\begin{equation}
t_{\rm dyn}(r)\equiv \sqrt{\frac{r^{3}}{G\,M_{\rm enc}(r)}},
\label{eq:tdyn_def}
\end{equation}
where $M_{\rm enc}(r)$ is the enclosed gravitating mass.
In the BH-dominated (``naked'') limit at $r\sim r_{\rm e}$, $M_{\rm enc}\approx M_{\rm BH}$, whereas in the soliton-dominated case $M_{\rm enc}\approx M_{\rm BH}+M_{\rm sol}(<r)\ge M_{\rm BH}$.
At fixed physical radius $r\sim r_{\rm e}\sim{\cal O}(r_c)$ this implies $t_{\rm dyn,sol}\le t_{\rm dyn,BH}$.
Since the instability is diagnosed by radiative losses/redistribution not being slow compared to dynamical support (i.e., $t_{\rm rad}\lesssim t_{\rm dyn}$), the conclusion that a long-lived hydrostatic atmosphere is untenable in the soliton case implies the same conclusion \emph{a fortiori} in the BH-only limit, where the dynamical time is longer.
The role of the soliton is therefore not to ``save'' a static atmosphere, but to define the characteristic LRD scale and to motivate considering a quasi-spherical, volume-filling configuration at $r\sim 50$--$100$ pc in the first place.
In what follows, we identify the characteristic physical extent of the obscurer/atmosphere directly with the observed effective radius $r_{\rm e}$, and assume $r_{\rm e}\sim {\cal O}(r_c)$.
\paragraph{A conservative column estimate from the baryon mass budget.}
Let a gas mass $M_g=f_g f_b M_s$ be loaded into an obscuring region of radius $r_{\rm e}=\xi r_c$, where $f_g\le 1$ is the loaded fraction of the cosmic baryon budget (we explore $f_g\sim 0.1$--$1$ to bracket feedback-driven depletion; e.g., \citealt{Hopkins2014}), $f_b\simeq 0.16$ \citep{Planck2020}, and $\xi\sim 1$--few.
A geometry-limited characteristic column is
\begin{equation}
N_{\rm H}\simeq \frac{M_g}{\pi r_{\rm e}^2\,\mu_{\rm H}m_p}
=\frac{f_g f_b M_s}{\pi (\xi r_c)^2\,\mu_{\rm H}m_p},\label{eq:NH_massbudget}
\end{equation}
with $\mu_{\rm H}\simeq 1.4$ accounting for He.
Using the soliton scaling (Eq.~(\ref{eq:rc_scaling})), this implies that for fixed $(f_g,\xi)$ the achievable column increases rapidly with deeper, more compact cores.
In Fig.~1 we define the ``Generic Transparency'' region by $N_{\rm H,max}(f_g=1,\xi=1)<10^{24}\,{\rm cm^{-2}}$.
As shown by the orange band in Fig.~1, adopting conservative values ($f_g=0.1, \xi=3$) shifts this threshold, but $m_{22}\approx 2$ remains a plausible solution for $M_s \sim 10^9 \msun$.
We note that Eq.~(\ref{eq:NH_massbudget}) is a conservative geometric estimate; gas clumping would increase the effective line-of-sight column, while feedback might reduce $f_g$.
\paragraph{Relation to hydrostatic stratification.}
For reference, an isothermal atmosphere in hydrostatic equilibrium satisfies $\nabla P=-\rho_g\nabla\Phi_s$, with
\begin{equation}
\rho_g(r)=\rho_{g,0}\exp\!\left[-\frac{\Phi_s(r)-\Phi_s(0)}{c_s^2}\right],\qquad c_s^2=\frac{k_B T}{\mu m_p}.
\end{equation}
This becomes a closed problem only after specifying a normalization.
Hydrostatic stratification can increase the central line-of-sight column relative to Eq.~(\ref{eq:NH_massbudget}); we therefore treat Eq.~(\ref{eq:NH_massbudget}) as a conservative baseline.
\paragraph{Cooling versus dynamical support.}
The mean density corresponding to a loaded core is $\bar n \simeq \bar\rho / \mu m_p$.
For the dense cores relevant here ($\bar n\gtrsim 10^{4}$--$10^{5}\,{\rm cm^{-3}}$), optically thin metal-line cooling at $T\sim 10^6$ K implies a short cooling time,
\begin{equation}
t_{\rm cool}\sim \frac{3k_BT}{2\bar n\,\Lambda(T,Z)},\label{eq:tcool}
\end{equation}
where $\Lambda(T,Z)$ is the radiative cooling function (we use the tabulations of \citealt{Sutherland1993}),
which can be much shorter than the potential-defined dynamical time $t_{\rm dyn}(r_{\rm e})$ (Eq.~\ref{eq:tdyn_def}).
For reference, a self-gravitating free-fall time would scale as $\sim (G\bar\rho)^{-1/2}$;
in the soliton case the relevant support time is set by the enclosed (BH+soliton) potential.
\paragraph{Radiative trapping: electron scattering vs.\ IR dust opacity.}
In an optically thick core, radiative losses may be limited by diffusion rather than by optically thin cooling.
A conservative lower bound is the diffusion time due to electron scattering, where $\tau_T\equiv\sigma_T N_H$ is the Thomson optical depth,
\begin{equation}
\begin{split}
t_{\rm diff,es} \sim & \frac{\tau_T r_{\rm e}}{c} = \frac{\sigma_T N_H r_{\rm e}}{c} \\
\simeq & 1.6\times 10^{3}\,{\rm yr}\,\left(\frac{\tau_T}{10}\right)\left(\frac{r_{\rm e}}{50\,{\rm pc}}\right),
\end{split}
\end{equation}
which is typically shorter than $t_{\rm dyn}$ in the parameter range where Compton-thick columns are attainable.
If the gas cools and becomes dusty, UV/optical photons can be efficiently reprocessed into the IR, and the relevant mass opacity can be much larger than the electron-scattering opacity $\kappa_{\rm es}\simeq 0.34~{\rm cm^2\,g^{-1}}$.
In that regime, an upper-range diffusion estimate is
\begin{equation}
t_{\rm diff,IR}\sim \frac{\tau_{\rm IR} r_{\rm e}}{c},\quad\tau_{\rm IR} = \kappa_{\rm IR}\Sigma,\quad\Sigma = \mu_H m_p N_H,
\end{equation}
so that
\begin{equation}
\begin{split}
t_{\rm diff,IR}\simeq & 3.8\times 10^{4}\,{\rm yr}\,\left(\frac{\kappa_{\rm IR}}{10\,{\rm cm^2\,g^{-1}}}\right) \\
& \times \left(\frac{N_H}{10^{25}\,{\rm cm^{-2}}}\right)\left(\frac{r_{\rm e}}{50\,{\rm pc}}\right).
\end{split}
\end{equation}
We stress that dust opacity is only relevant once the gas has cooled to temperatures where dust can survive;
in an initially hot ($T\sim T_{\rm vir}$) phase, electron scattering provides the appropriate minimal trapping floor.
\paragraph{Working definition of the ``Opacity Crisis.''}
We use ``Opacity Crisis'' to denote the breakdown of a long-lived, hot hydrostatic atmosphere inside the soliton: 
in the same parameter space where $N_H\gtrsim 10^{24}\,{\rm cm^{-2}}$ is achievable within $r_{\rm e}\sim O(r_c)$, the effective radiative-loss or energy-redistribution time $t_{\rm rad}\equiv \max(t_{\rm cool},t_{\rm diff,es})$ is not $\gg t_{\rm dyn}$.
The system therefore cannot remain as a quasi-static hot atmosphere;
it must evolve dynamically, either via rapid cooling-driven collapse ($t_{\rm cool}\ll t_{\rm dyn}$) or into a radiation-pressure-supported optically thick inflow/envelope if IR trapping becomes important ($t_{\rm diff,IR}\gtrsim t_{\rm dyn}$).
In either case, the initial static HSE picture is invalid in the regime relevant for LRD-like obscuration.
\subsection{Timescale analysis and inflow rate}
\begin{figure}[ht!]
\centering
\safeincludegraphics[width=\columnwidth]{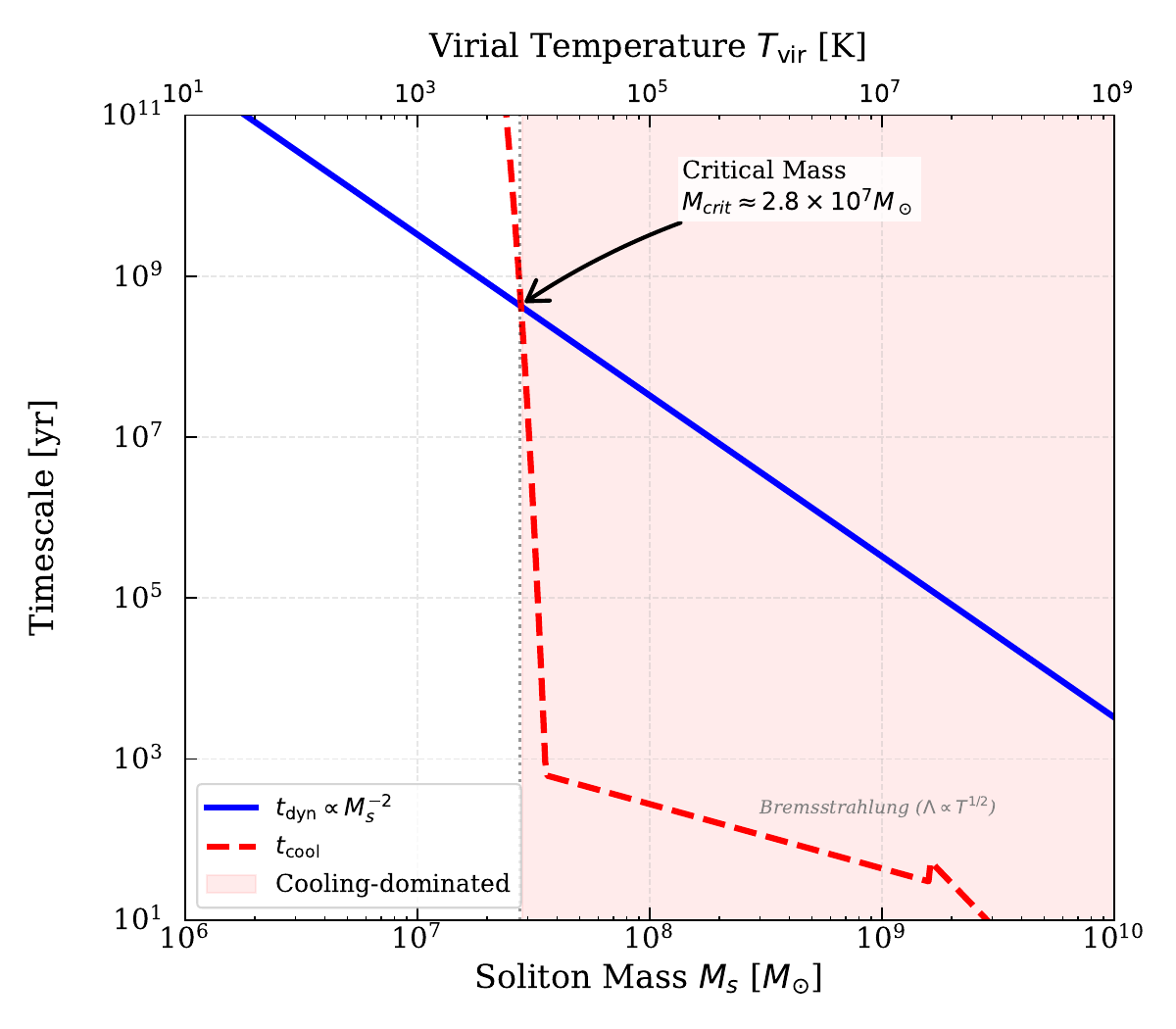}
\caption{Timescale hierarchy in soliton cores. Comparison between the
dynamical time $t_{\rm dyn}$ (blue solid line, $t_{\rm dyn} \propto M_s^{-2}$)
and the optically thin cooling time $t_{\rm cool}$ (red dashed line;
this corresponds strictly to the optically thin cooling time defined
in Equation (5), not the effective diffusion-limited time $t_{\rm rad}$)
as a function of soliton mass $M_s$. The top axis explicitly provides
the corresponding virial temperature $T_{\rm vir}$ to guide physical
intuition. At low masses ($T_{\rm vir} < 10^4\,{\rm K}$), the exponential
suppression of atomic cooling in primordial gas leads to a stable regime
where $t_{\rm cool} > t_{\rm dyn}$. As the soliton mass increases, the gas
is heated above the atomic cooling threshold, causing a sharp drop in
$t_{\rm cool}$. A critical mass emerges at $M_{\rm crit} \approx
2.8 \times 10^7 M_\odot$, identified exactly at the intersection where
the condition $t_{\rm cool} = t_{\rm dyn}$ is satisfied; this threshold
is obtained analytically by solving $t_{\rm cool} = t_{\rm dyn}$ using
the cooling function adopted in our model (see Appendix B). Above this
mass threshold, the system enters the cooling-dominated regime
($t_{\rm cool} < t_{\rm dyn}$). For the massive regime relevant to Little
Red Dots ($M_s \gtrsim 10^8 M_\odot$), the cooling time is orders of
magnitude shorter than the dynamical time, inevitably triggering a
catastrophic collapse.}
\label{fig:timescales}
\end{figure}

Figure~\ref{fig:timescales} summarizes the timescale hierarchy for our fiducial choice $m_{22}=2.0$.
In the parameter space where Eq.~(\ref{eq:NH_massbudget}) allows $N_{\rm H}\gtrsim 10^{24}\,{\rm cm^{-2}}$, radiative losses are rapid compared to dynamical support.
A direct implication is that the effective sound speed can drop substantially during collapse, enhancing the capture rate.
For a black hole of mass $M_{\rm BH}$, the Bondi rate is
\begin{equation}
\dot M_{\rm B}=4\pi \lambda \frac{G^2 M_{\rm BH}^2\rho}{c_s^3}.
\end{equation}
Here $\lambda$ is the usual dimensionless Bondi factor (order unity, depending on the equation of state), and $\eta$ is the radiative efficiency used in defining $\dot M_{\rm Edd}$.
The ratio to the Eddington rate is
\begin{equation}
\begin{split}
\frac{\dot M_{\rm B}}{\dot M_{\rm Edd}}\approx & 10^3\,\left(\frac{\lambda}{1}\right)\left(\frac{\eta}{0.1}\right)\left(\frac{M_{\rm BH}}{10^8 M_\odot}\right) \\
& \times \left(\frac{n}{10^5\,{\rm cm^{-3}}}\right)\left(\frac{T}{10^6\,{\rm K}}\right)^{-3/2},
\end{split}
\label{eq:bondi_ratio}
\end{equation}
indicating that inflow at the capture scale can exceed $\dot M_{\rm Edd}$ in dense cores.
In the soliton scenario, the compact potential well both confines baryons within $r\sim r_c$ (raising $n$ for a given loaded mass) and sets the dynamical boundary condition for collapse, so that once cooling reduces $c_s$ the Bondi rate enhancement in Eq.~(\ref{eq:bondi_ratio}) is naturally realized at the LRD scale;
in a more weakly bound BH-only or extended-halo configuration the same column would correspond to a less compact reservoir and a smaller capture enhancement.
Granule-driven fluctuations (Quantum Bondi Boost) may further modulate this \citep{Chiu2025}.

\subsection{Thermodynamic Instability and the Cooling Catastrophe}
\label{sec:instability_derivation}

Figure~\ref{fig:timescales} summarizes the hierarchy between the dynamical timescale $t_{\rm dyn}$ and the radiative cooling timescale $t_{\rm cool}$ within the soliton core.
Here we provide a scaling argument for why a crossover $t_{\rm cool}<t_{\rm dyn}$ is unavoidable at sufficiently large soliton masses.
\paragraph{Potential-defined dynamical time.}
The relevant dynamical timescale at radius $r$ is set by the enclosed gravitating mass,
\begin{equation}
t_{\rm dyn}(r)\equiv \sqrt{\frac{r^{3}}{G\,M_{\rm enc}(r)}}.
\end{equation}
Evaluated at the core scale ($r\sim r_c$) in the soliton-dominated regime, we estimate
\begin{equation}
t_{\rm dyn}(r_c)\sim \sqrt{\frac{r_c^{3}}{G\,M_{\rm enc}(r_c)}}\;\simeq\;\sqrt{\frac{r_c^{3}}{G\,[M_{\rm BH}+{\cal O}(1)M_s]}}.
\label{eq:tdyn_rc}
\end{equation}
Using the soliton scaling $r_c\propto m_{22}^{-2}M_s^{-1}$ \citep{Schive2014a}, for fixed particle mass $m_{22}$ this implies $t_{\rm dyn}\propto M_s^{-2}$ (and $t_{\rm dyn}\propto m_{22}^{-3}$ at fixed $M_s$).
\paragraph{Cooling time for a hydrostatic atmosphere.}
If a gas mass $M_g=f_g f_b M_s$ is distributed within $r_{\rm e}=\xi r_c$, the characteristic number density scales as
\begin{equation}
n \propto \frac{M_g}{r_{\rm e}^{3}} \propto \frac{f_g M_s}{(\xi r_c)^3}\propto f_g\,\xi^{-3}\,M_s^{4},
\end{equation}
and the characteristic temperature for a hydrostatic atmosphere is set by the potential depth,
\begin{equation}
T\sim T_{\rm vir}\propto \frac{M_{\rm enc}}{r_c}\propto M_s^{2}
\quad\text{(soliton-dominated regime)}.
\end{equation}
The cooling time is $t_{\rm cool}\sim (n k_B T)/(n^2\Lambda)$.
In the high-temperature regime where thermal bremsstrahlung is a good scaling proxy ($\Lambda\propto T^{1/2}$), this gives
\begin{equation}
t_{\rm cool}\propto \frac{T^{1/2}}{n}\propto M_s^{-3}.
\label{eq:tcool_scaling}
\end{equation}
Figure~\ref{fig:timescales} uses the full cooling function for the numerical estimate; the scalings above are provided for intuition.
\paragraph{Inevitable crossover and the role of the soliton.}
In primordial gas lacking molecular or metal coolants, radiative cooling is exponentially suppressed below $T \sim 10^4$ K due to the Boltzmann factor associated with hydrogen collisional excitation (Lyman-$\alpha$ cooling).
Consequently, low-mass solitons ($M_s \lesssim 10^7 M_\odot$, $T_{\rm vir} < 10^4$ K) can maintain a stable, quasi-static atmosphere where the cooling time strictly exceeds the dynamical time ($t_{\rm cool} > t_{\rm dyn}$).
However, as the soliton mass increases, the deeper potential well heats the gas above the $10^4$ K threshold, triggering highly efficient atomic cooling.
In the high-temperature regime, $t_{\rm cool}$ decreases more steeply with soliton mass (e.g., $\propto M_s^{-3}$ for bremsstrahlung) than $t_{\rm dyn}$ ($\propto M_s^{-2}$).
This guarantees a crossover where $t_{\rm cool} < t_{\rm dyn}$.
As illustrated in the updated Figure~\ref{fig:timescales}, this transition defines a critical mass scale at $M_{\rm crit} \approx 2.8 \times 10^7 M_\odot$.
The massive FDM cores relevant to LRDs ($M_s \gtrsim 10^8 M_\odot$) already reside deeply within this cooling-dominated region.
In this unstable regime, the gas loses thermal support faster than the potential-defined dynamical time, triggering rapid inflow or collapse rather than maintaining a long-lived hot atmosphere.
For comparison, at the same physical radius $r\sim r_c$, a BH-only (``naked'') potential has $M_{\rm enc}\approx M_{\rm BH}$ and therefore a \emph{longer} dynamical time than the soliton case (because $M_{\rm enc,sol}\ge M_{\rm BH}$).
Thus, if $t_{\rm cool}\lesssim t_{\rm dyn}$ holds in the soliton container, it holds \emph{a fortiori} in the BH-only limit.
The role of the soliton is therefore not to stabilize the gas, but to define the characteristic $\sim 50$--$100$ pc scale relevant to LRDs and to motivate considering a quasi-spherical, volume-filling atmosphere at that scale.
\section{Radiative Implications}\label{sec:radiative}

\subsection{X-ray obscuration}
The primary observational puzzle of LRDs is their X-ray faintness.
Using the conservative estimate in Eq.~(\ref{eq:NH_massbudget}), typical columns can reach $N_{\rm H}\sim 10^{24}$--$10^{25}\,{\rm cm^{-2}}$.
The Thomson optical depth is $\tau_T=\sigma_T N_{\rm H}\simeq 0.67\,(N_{\rm H}/10^{24}\,{\rm cm^{-2}}).$
For $\tau_T\gtrsim 1$, the direct soft X-ray continuum is strongly suppressed.
We expect the observed X-ray signal to be dominated by a scattered fraction $f_{\rm scat}$.
For typical AGN geometries, $f_{\rm scat} \sim 1-5\%$, implying a suppression of the apparent $L_X$ by $\sim 1.5-2$ dex, consistent with \citet{Maiolino2024}.
\subsection{Dust Attenuation and Broad Line Visibility}
Recent analyses suggest that LRDs are best described as young SMBHs embedded in compact, physically thick obscurers (``dense ionized cocoons'' and/or heavy dust shrouds) with column densities $\nh \gtrsim 10^{24}\,{\rm cm^{-2}}$ \citep[e.g.,][]{Rusakov2026}.
Our FDM soliton framework provides a natural origin for such a geometry: unlike a rotationally supported thin disk, the soliton produces a quasi-spherical potential well on the characteristic scale $r_c\sim 50$--$100$ pc, so gas inflowing into the core can initially fill a large fraction of the volume.
\paragraph{A simple parameterized cocoon geometry.}
We parameterize the obscurer as extending from an inner dust sublimation scale $r_{\rm in}$ to an observable effective radius
\begin{equation}
r_{\rm e} \simeq \xi\,r_c,
\end{equation}
with a line-of-sight covering factor $f_{\rm cov}\gtrsim 0.5$.
A convenient phenomenological description is a biconical polar opening with half-opening angle $\theta$, for which $f_{\rm cov}\simeq \cos\theta$.
In this picture, broad lines remain visible either through low-column sightlines (gaps) or via scattering into the line of sight, while the high covering factor naturally accounts for the high detection rate of obscured LRDs \citep[e.g.,][]{Greene2024}.
\paragraph{Granularity and clumpiness in the soliton core.}
A specific ingredient of the FDM core is its intrinsic time-dependent granularity arising from wave interference.
Order-unity density fluctuations (``granules''; $\delta\rho/\rho\sim 1$) produce persistent gravitational stirring on timescales comparable to the soliton dynamical/oscillation time.
These perturbations can \emph{promote} and help maintain a clumpy cocoon/torus morphology, providing a complementary route to clumpiness beyond magnetic or radiation-driven instabilities.
\paragraph{How cocoon observables can constrain the FDM scenario.}
In this framework the ``Opacity Crisis'' describes the global collapse/evolution of a (possibly clumpy) atmosphere once $t_{\rm rad}\lesssim t_{\rm dyn}$.
Over time, feedback may carve polar channels and reduce $f_{\rm cov}$, but the deeper potential of a soliton can qualitatively help retain a high-column equatorial component longer than a shallow, baryon-dominated halo.
Observables such as the broad-line visibility fraction, polarization, and inferred covering factors/opening angles (and their evolution with mass and redshift) therefore provide potential empirical constraints on the depth of the central potential and, by extension, the allowed range of $m_{22}$ \citep[see also][]{Delvecchio2025}.
\subsection{Thermodynamic Equilibrium}
Assuming the obscured luminosity is re-radiated in the IR, a characteristic effective temperature follows from the Stefan--Boltzmann law.
For $L_{\rm bol}\sim10^{46}\,{\rm erg\,s^{-1}}$ and a photospheric scale $R_{\rm ph}\sim{\cal O}(r_c)$, we obtain $T_{\rm eff}\approx150$--$200\,{\rm K}$, implying a rest-frame mid-IR bump (peak near $\sim 20\,\mu{\rm m}$).
The very red JWST colors at shorter wavelengths likely require an additional hotter dust component (near the sublimation radius) and/or anisotropic leakage/scattering through a non-uniform cocoon;
quantifying the detailed SED requires radiative-transfer/radiation-hydrodynamic modeling. In this context the soliton potential primarily sets the outer scale and confinement (through $r_c$ and $r_{\rm e}$);
it does not by itself determine the dust microphysics at $r_{\rm in}$, so the need for a hot dust component and its dependence on halo/core properties ultimately requires dedicated radiation-hydrodynamic simulations.
\section{Numerical Methodology}\label{sec:method}

Our analytic picture requires that deep, compact solitonic potentials form robustly. We validate this using 3D simulations.
\subsection{Setup and Initial Conditions}
We solve the Schr\"odinger--Poisson system using a pseudo-spectral code on a $N=512^3$ grid with periodic boundary conditions \citep[see][]{Schive2014a}.
Our aim is not to claim the discovery of soliton formation, but to demonstrate that compact cores with LRD-like scales form robustly in this setup.
To model the formation of the potential well, we initialize the wavefunction with eight identical ground-state solitons with zero initial phases (free-fall merger).
While we use zero phase for demonstration, previous studies have shown soliton formation is robust against random phase fluctuations \citep{Schive2014b}.
The simulations were executed on an NVIDIA RTX 3080, requiring approximately one day of runtime.

\subsection{Physical 
Scaling}
For $\mtwotwo = 2.0$, we scale our simulation box to $L_{\rm phys} = 2.0$ kpc.
This maps our initial configuration to 8 solitons of mass $M_{\rm ini} \approx 3.9 \times 10^8 \msun$ each.
The final central soliton contains $\Mc \approx 7.8 \times 10^8 \msun$ with a radius $\rc \approx 51$ pc.
This creates a representative scaling consistent with LRDs.

\section{Simulation Results}\label{sec:results}

\begin{figure}[t]
    \centering
    \safeincludegraphics[width=\columnwidth]{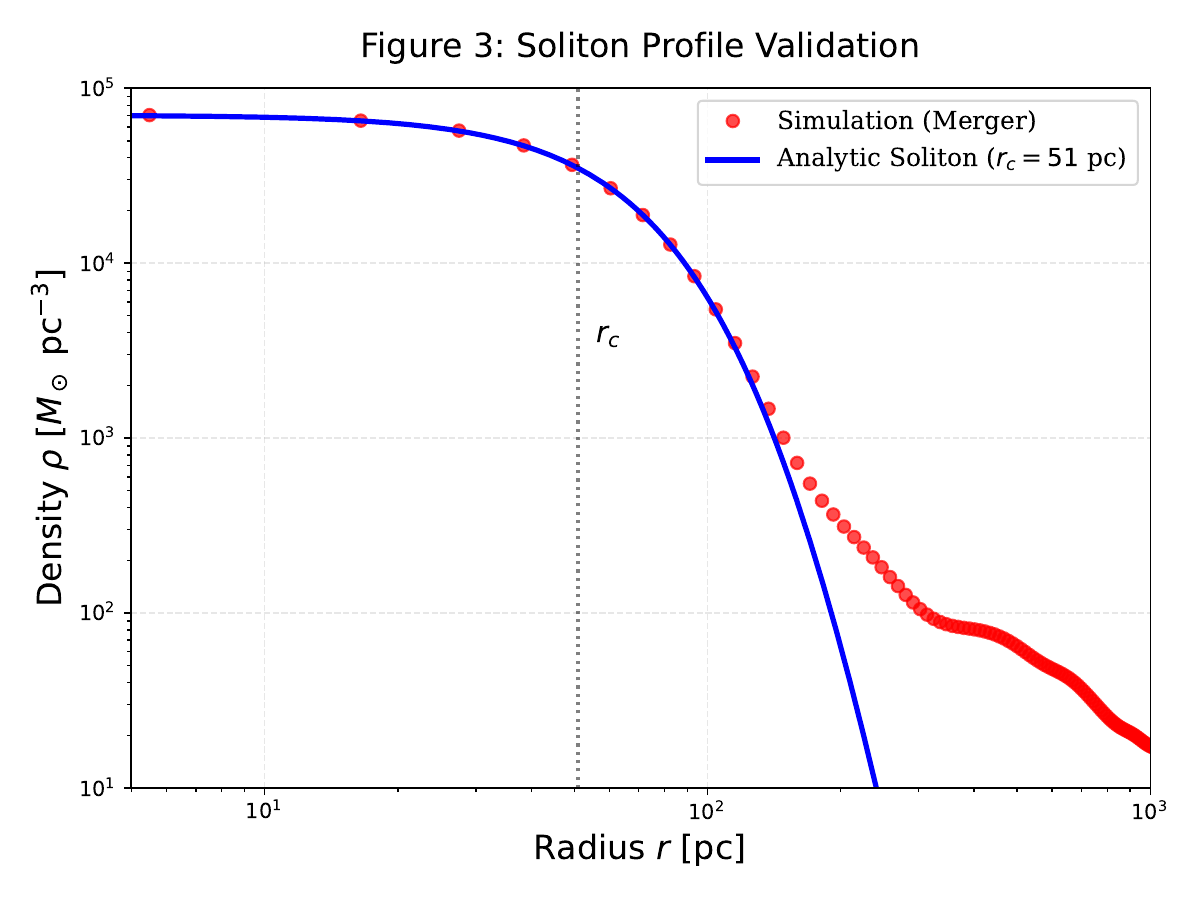}
    \caption{Validation of a compact soliton core in an idealized merger.
Radial density profile of the simulated core (red points) compared to the analytic soliton solution (black line; fitted $r_c=2.32$ in code units).
Both axes are shown in code units; the corresponding physical scaling for $\mtwotwo=2$ is given in Section~\ref{sec:method} (yielding $r_c\simeq 51$~pc and $M_c\simeq 7.8\times 10^8\,M_\odot$).
The deviation at large radii reflects unrelaxed merger debris and finite-box/periodic-boundary effects, and does not affect the inner soliton core relevant for our analytic estimates.}
    \label{fig:profile}
\end{figure}

Figure~\ref{fig:profile} shows the density profile of the final core.
The inner region is well described by the soliton profile, illustrating that violent relaxation in idealized mergers can form a compact solitonic potential well.
Under our representative physical scaling, the core has $M_c\sim 8\times 10^8\,M_\odot$ and $r_c\sim 50$ pc, placing it in the range considered in our analytic estimates.

\section{Discussion}\label{sec:discussion}

\subsection{Resolution of the Lyman-$\alpha$ Tension}
\label{sec:tension}

\begin{figure}[t]
    \vspace{15pt}
    \centering
    \safeincludegraphics[width=\columnwidth]{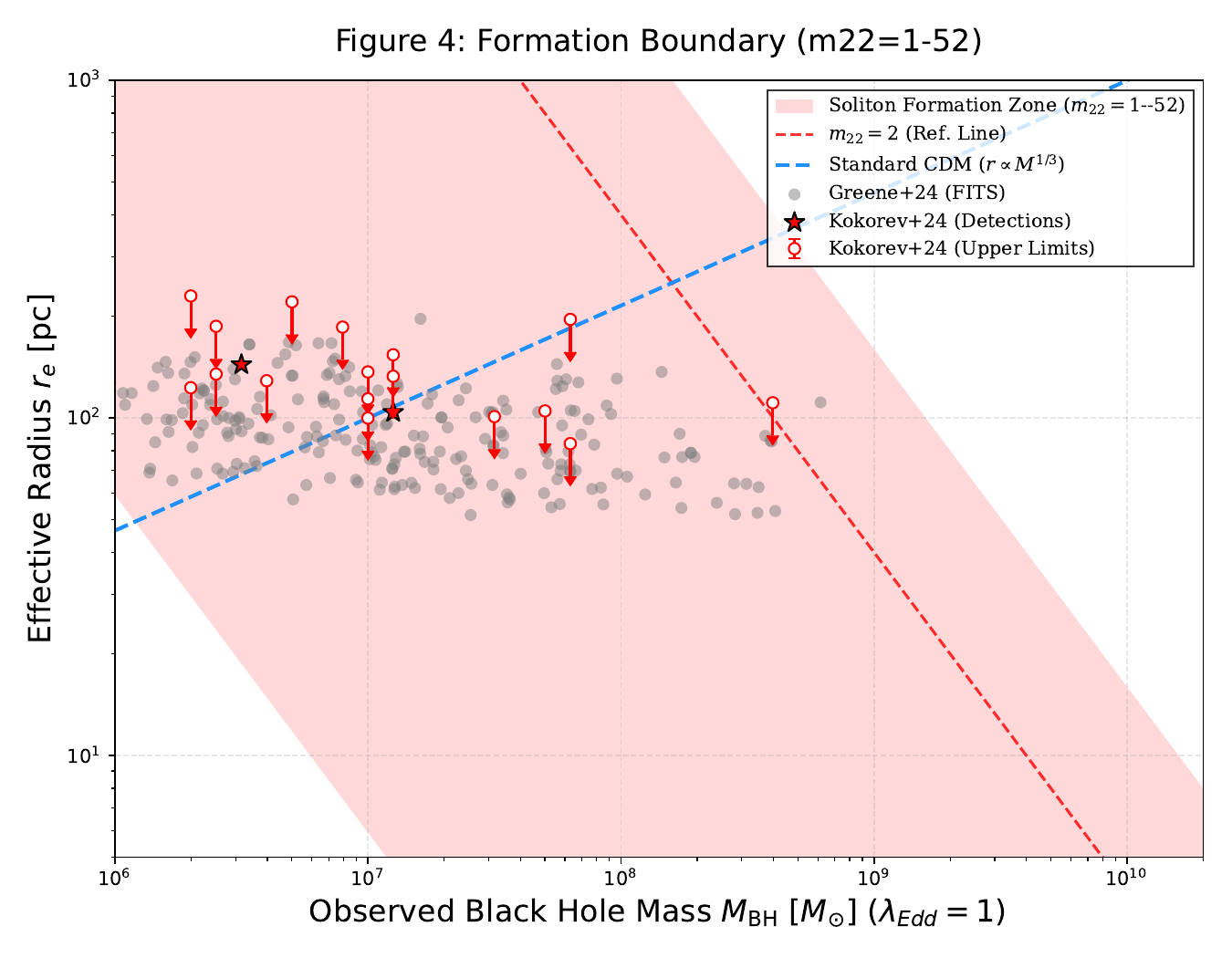}
    \caption{\textbf{The size--mass plane of Little Red Dots (LRDs), comparing observed effective radii $r_e$ with illustrative FDM soliton compactness limits.} The red shaded region indicates a ``Soliton Formation Zone,'' defined by the minimum radii implied by the soliton scaling for different boson masses.
The horizontal axis represents the \textbf{observed black hole mass ($M_{\rm BH}$)}, which does not necessarily equal the underlying soliton mass ($M_c$) during the early progenitor phase.
The dashed red line shows our fiducial baseline ($m_{22}=2$) under the direct-mapping assumption ($M_{\rm BH} \approx M_c$).
LRD measurements from \citet{Greene2024} and \citet{Kokorev2024} are overplotted. In the progenitor phase interpretation ($M_{\rm BH} < M_c$), the horizontal scatter of the observational data reflects the growth track of black holes at a fixed $m_{22}$ within larger, fixed soliton cores ($M_c \gg M_{\rm BH}$), rather than a varying $m_{22}$.
This resolves the apparent tension with the $m_{22} \approx 2$ estimate in Section~\ref{sec:theory}.1, which strictly assumes a mature, direct-mapping ($M_{\rm BH} \approx M_c$) scenario. This interpretation allows the observed LRDs to be consistent with heavier boson masses ($m_{22} \gtrsim 20$, consistent with Lyman-$\alpha$ constraints) while still utilizing the deep soliton potential to trigger the initial collapse.}
    \label{fig:size_mass}
    \vspace{15pt}
\end{figure}

While a fiducial boson mass of $m_{22} \approx 2$ allows for a direct mapping where the black hole mass tracks the soliton mass ($M_{\rm BH} \approx M_c$), our analysis of the size--mass relation (Figure~\ref{fig:size_mass}) reveals that the observed data are fully compatible with higher particle masses (e.g., $m_{22} \sim 20$--$50$, consistent with Lyman-$\alpha$ forest constraints; see, e.g., \citealt{Irsic2017, RogersPeiris2021}) if LRDs represent an \textbf{early progenitor phase} where the 
central black hole has not yet consumed the entire soliton core.
We note that the precise Lyman-$\alpha$ bound depends on assumptions about the IGM thermal history and on whether FDM constitutes all of the dark matter;
mixed-DM scenarios with a subdominant FDM fraction can weaken the constraint \citep[e.g.,][]{Kobayashi2017}.
In this scenario, the total soliton mass $M_c$ remains larger than the growing black hole mass $M_{\rm BH}$.
By relaxing the strict assumption of $M_{\rm BH} = M_c$ and interpreting the horizontal scatter in Figure~\ref{fig:size_mass} (where the axis represents observed $M_{\rm BH}$) as a reflection of evolutionary stages ($M_{\rm BH} < M_c$), the parameter space naturally opens up.
To clarify the apparent tension with Section~\ref{sec:theory}.1: the $m_{22} \approx 2$ baseline is strictly derived under a mature, direct-mapping assumption ($M_{\rm BH} \approx M_c$).
However, during the formation stage, the host soliton mass can be substantially larger than the growing black hole ($M_c \gg M_{\rm BH}$).
Therefore, the horizontal distribution of the observational data reflects black hole growth (increasing $M_{\rm BH}$) within larger, fixed soliton cores (fixed $m_{22}$), rather than varying $m_{22}$ at a fixed evolutionary stage.
Under this progenitor interpretation, $m_{22} \gtrsim 20$ is fully consistent with both LRD sizes and Lyman-$\alpha$ limits, without contradicting the fiducial mature-phase estimate.
The soliton provides the deep, compact potential well required to trigger the opacity crisis, while the observed black hole mass reflects only the collapsed fraction.
Consequently, the apparent tension between the compact sizes of LRDs and the Lyman-$\alpha$ constraints may be substantially alleviated, suggesting that the FDM framework remains compatible with both the observed small-scale structure of LRDs and large-scale Lyman-$\alpha$ forest data.
\subsection{Observational Signatures}\label{sec:predictions}

We provide specific observational predictions:

\begin{itemize}
    \item \textbf{Inverse Size--Mass Relation:} For soliton cores at fixed $m_{22}$, the soliton scaling implies $r_c\propto M_c^{-1}$.
Figure~\ref{fig:size_mass} illustrates this compactness floor and the observed population above it.
Establishing an intrinsic inverse $r_c$--$M_c$ relation as a discriminator will require independent constraints on host/core masses (beyond virial $M_{\rm BH}$ estimates) and careful modeling of how radiative transfer maps the intrinsic core scale to the observed $r_e$.
\item \textbf{Polarization:} 
    Our clumpy dust cocoon model inevitably imprints a polarization signal on the broad emission lines \citep[e.g.,][]{Greene2024}.
\item \textbf{Compact Hot Dust vs. ALMA Non-detections:} 
    Recent ALMA observations constrain the cold dust mass ($M_{dust} \lesssim 10^6 \msun$; \citealt{Casey2025}).
The ``Opacity Crisis'' creates extreme gas column densities within a tiny volume, consistent with low total dust mass despite high optical depth.
\end{itemize}

\subsection{Demographics and Duty Cycle}
A key question is whether the transient obscured phase implied by the ``Opacity Crisis'' can match the
observed abundance of LRDs, $n_{\rm LRD}\sim 10^{-5}\,{\rm comoving\ Mpc^{-3}}$ (here cMpc denotes comoving Mpc).
For massive halos at $z\sim 5$--$7$, a representative number density is $n_{\rm halo}\sim 10^{-4}\,{\rm cMpc^{-3}}$ \citep[e.g.,][]{Kokorev2024},
suggesting an instantaneous occupancy fraction of order $f_{\rm occ}\equiv n_{\rm LRD}/n_{\rm halo}\sim 0.1$.
More generally, the occupancy can be written as a time-averaged duty cycle,
\begin{equation}
n_{\rm LRD}\;\simeq\;
n_{\rm halo}\,P_{\rm trig}\,N_{\rm evt}\left(\frac{t_{\rm evt}}{t_H}\right),
\end{equation}
where $P_{\rm trig}$ is the probability that a given halo experiences an LRD-triggering event within the relevant
redshift window, $N_{\rm evt}$ is the number of such episodes per halo, $t_{\rm evt}$ is the observable duration of each episode,
and $t_H$ is the Hubble time at that epoch.
While the characteristic dynamical time of the dense core can be short ($t_{\rm dyn}\sim \sqrt{r_c^3/(G M_{\rm enc})}\sim 10^{5}\,{\rm yr}$ for $r_c\sim 50$~pc and $M_{\rm enc}\sim 10^9\,M_\odot$), feedback, clumpy obscuration,
and/or repeated episodes can extend the effective observable duration to $t_{\rm evt}\sim 10^{6}$--$10^{7}\,{\rm yr}$.
Matching $f_{\rm occ}\sim 0.1$ then requires $P_{\rm trig}N_{\rm evt}$ to be of order unity to several, motivating multiple
merger-driven or inflow-driven obscured episodes during early assembly.

\section{Conclusion}\label{sec:conclusion}
We have presented a compact model in which LRD-like systems arise during a transient, heavily obscured phase inside FDM soliton cores.
Combining size scaling with opacity constraints suggests $m_{22}\sim{\rm few}$. A conservative column estimate, including sensitivity analysis for baryon loading, indicates that reaching $N_{\rm H}\gtrsim 10^{24}\,{\rm cm^{-2}}$ typically implies radiative losses faster than dynamical support (or radiation-pressure dominance), disfavoring a long-lived static hot atmosphere and motivating rapid inflow or radiation-pressure-driven evolution.
In progenitor-phase interpretations where $M_{\rm BH} < M_c$, the observed LRD population may also remain consistent with heavier boson masses ($m_{22} \gtrsim 20$), while still triggering the same opacity-crisis collapse mechanism.
\appendix
\setcounter{equation}{0}
\section*{Supplementary Notes: Detailed Derivations and Physical Interpretations}

This document provides rigorous mathematical derivations for the scaling laws and physical timescales presented in the revised manuscript.
Specifically, we explicitly derive the dynamical timescale based on the soliton gravitational potential and compare it with the radiative cooling timescale.
We demonstrate that the inclusion of the soliton potential significantly deepens the central well compared to a naked black hole;
however, even with this enhanced gravitational support, radiative losses dominate for massive cores, leading to the unavoidable ``Opacity Crisis.''

\section{Soliton Scaling Laws (Section 2.1)}

\subsection{Equation 1: The Soliton Core Radius}
\textbf{Equation:}
\begin{equation}
r_{c} \approx 160 \, \text{pc} \, m_{22}^{-2} \left(\frac{M_{s}}{10^{9}M_{\odot}}\right)^{-1}
\end{equation}

\noindent \textbf{Derivation:} 
The Fuzzy Dark Matter (FDM) soliton represents the ground-state solution of the Schrödinger-Poisson system.
The scale arises from the balance between quantum pressure (uncertainty principle) and self-gravity.
The de Broglie wavelength is $\lambda_{dB} \sim \frac{\hbar}{mv}$. For a self-gravitating system, the virial velocity is $v \sim \sqrt{GM/r}$.
Equating the quantum pressure scale $r \sim \lambda_{dB}$ with the gravitational scale:
\begin{equation}
r \sim \frac{\hbar}{m\sqrt{GM/r}} \implies r^{1/2} \sim \frac{\hbar}{m\sqrt{GM}} \implies r \sim \frac{\hbar^2}{G m^2 M}
\end{equation}
Substituting physical constants yields the inverse scaling $r_{c} \propto m^{-2} M_{s}^{-1}$.
\section{Timescale Hierarchy and Stability (Sections 2.2 \& 2.4)}

\subsection{Dynamical Timescale: Soliton vs. Naked Black Hole}
\textbf{Definition:} 
We define the dynamical time at the core radius $r_c$ based on the gravitational potential of the \textit{total enclosed mass}.
This distinguishes the ``Soliton Container'' scenario from a naked black hole scenario.
\noindent \textbf{Derivation:} 
The dynamical time is defined as:
\begin{equation}
t_{dyn}(r) \equiv \sqrt{\frac{r^3}{G M_{enc}(r)}}
\end{equation}

\vspace{15pt}
\noindent \textbf{Case A: Soliton-Dominated Potential (The LRD Scenario)} \\
For a black hole embedded in a soliton, $M_{enc}(r_c) \approx M_{BH} + M_{s}(<r_c)$.
In the progenitor phase or massive soliton limit where $M_s \gtrsim M_{BH}$, the potential is dominated by the soliton mass $M_s$.
Substituting the soliton scaling relation $r_{c} \propto m^{-2} M_{s}^{-1}$:
\begin{equation}
t_{dyn, sol} \propto \sqrt{\frac{(m^{-2}M_{s}^{-1})^3}{M_{s}}} = \sqrt{m^{-6} M_{s}^{-4}} = m^{-3} M_{s}^{-2}
\end{equation}
Thus, for a fixed boson mass $m$, the dynamical support time scales as:
\begin{equation}
t_{dyn} \propto M_{s}^{-2}
\end{equation}
This indicates that more massive solitons have quadratically shorter dynamical times due to their deeper and more compact potentials.
\vspace{15pt}

\noindent \textbf{Case B: Naked Black Hole (Comparison for Major Point 1)} \\
In the absence of a soliton (a ``naked'' black hole), the enclosed mass at the same radius $r_c$ is simply $M_{enc} \approx M_{BH}$.
Since $M_{enc, sol} > M_{enc, BH}$ (often significantly, as $M_s \sim 10^9 M_\odot$ while $M_{BH}$ may be growing), we have:
\begin{equation}
t_{dyn, sol} < t_{dyn, BH}
\end{equation}
The soliton creates a deeper potential well, providing \textit{stronger} gravitational support (shorter collapse time) than the black hole alone.
The fact that the ``Opacity Crisis'' (instability) occurs even in the soliton case ($t_{cool} < t_{dyn, sol}$) implies that a static atmosphere is \textit{a fortiori} impossible in the naked black hole case, where the dynamical support is weaker (longer $t_{dyn}$).
The role of the soliton is therefore to define the compact physical scale ($r_c \sim 50$ pc) and density threshold where this crisis manifests.
\subsection{Cooling Timescale Derivation}
\textbf{Gas Properties:} 
We assume the gas initially traces the soliton potential in hydrostatic equilibrium.
\begin{itemize}
    \item \textbf{Virial Temperature:} $T \propto \frac{M_s}{r_c} \propto \frac{M_s}{M_s^{-1}} \propto M_s^2$.
\item \textbf{Gas Density:} Assuming gas mass $M_g \propto M_s$ fills the volume $r_c^3$:
    \begin{equation}
    n \approx \frac{M_g}{m_p r_c^3} \propto \frac{M_s}{(M_s^{-1})^3} \propto M_s^4
    \end{equation}
\end{itemize}

\noindent \textbf{Cooling Time and Multiphase Scaling:} 
The cooling timescale is given by $t_{\rm cool} \sim \frac{k_B T}{n \Lambda(T)}$.
Given $n \propto M_s^4$ and $T \propto M_s^2$, the scaling depends on the cooling function $\Lambda(T)$:

\begin{itemize}
    \item \textbf{Atomic Cooling Threshold ($T < 10^4$ K):} For $M_s < 2.8 \times 10^7 M_\odot$, $\Lambda(T)$ is exponentially suppressed by the Boltzmann factor.
Consequently, $t_{\rm cool}$ increases sharply as $M_s$ decreases, exceeding the dynamical time and creating a stable regime.
\item \textbf{Line Cooling Regime ($10^4 < T < 10^7$ K):} As implemented in our numerical model, $\Lambda(T) \propto T^{-0.6}$ in this interval.
The scaling becomes:
    \begin{equation}
    t_{\rm cool} \propto \frac{T}{n \Lambda(T)} \propto \frac{T^{1.6}}{n} \propto \frac{(M_s^2)^{1.6}}{M_s^4} = M_s^{-0.8}
    \end{equation}
    This explains the shallower slope of the cooling curve in the middle section of Figure 2.
    
    \item \textbf{Bremsstrahlung Regime ($T > 10^7$ K):} For massive cores ($M_{s} \sim 10^9 M_{\odot}$), thermal bremsstrahlung dominates, so $\Lambda(T) \propto T^{1/2}$.
Substituting the scalings yields:
    \begin{equation}
    t_{\rm cool} \propto \frac{T^{1/2}}{n} \propto \frac{(M_s^2)^{1/2}}{M_s^4} = M_s^{-3}
    \end{equation}
    This confirms the steep slope $t_{\rm cool} \propto M_s^{-3}$ shown in Figure 2 for the high-mass regime.
\end{itemize}

\subsection{The Crossover (The Opacity Crisis)}
Comparing the slopes derived above against the dynamical support:
\begin{itemize}
    \item Dynamical Support: $t_{\rm dyn} \propto M_{s}^{-2}$
    \item Radiative Cooling: $t_{\rm cool} \propto M_{s}^{-0.8}$ (intermediate) and $t_{\rm cool} \propto M_{s}^{-3}$ (high-mass)
\end{itemize}
The intersection $t_{\rm dyn} = t_{\rm cool}$ defines the stability limit.
Numerically solving for our fiducial parameters ($m_{22}=2.0$), we find a critical mass scale:
\begin{equation}
M_{\rm crit} \approx 2.8 \times 10^7 M_\odot
\end{equation}
For $M_s > M_{\rm crit}$, the gas loses thermal energy faster than the potential can dynamically readjust ($t_{\rm cool} < t_{\rm dyn}$).
This leads to catastrophic collapse rather than a static hot atmosphere, confirming the ``Opacity Crisis''.
\section{Demographics (Section 6.3)}

\subsection{Equation 17: Occupancy Fraction}
\textbf{Equation:}
\begin{equation}
n_{LRD} \simeq n_{halo} P_{trig} N_{evt} \left(\frac{t_{evt}}{t_{H}}\right)
\end{equation}
\textbf{Derivation:} 
This is a standard duty-cycle argument based on the Ergodic Hypothesis.
\begin{itemize}
    \item $n_{halo}$: Number density of potential host halos.
\item $P_{trig}$: Probability that a halo is triggered to host an LRD.
    \item $N_{evt}$: Number of events per triggered halo.
\item $t_{evt}/t_H$: The fractional duration of the event relative to the Hubble time.
\end{itemize}
The observed number density is simply the total halo density weighted by the probability of catching a halo in the active, obscured LRD phase.

\end{document}